\documentclass[preprint,aps,showpacs,preprintnumbers,amsmath,amssymb]{revtex4}

\usepackage{graphicx}
\usepackage{dcolumn}
\usepackage{bm}

\begin{document}


\title{NIR Femtosecond Control of Resonance-Mediated Generation of Coherent Broadband UV Emission}
\author{Leonid Rybak, Lev Chuntonov, Andrey Gandman, Naser Shakour}
\author{Zohar Amitay}
\email{amitayz@tx.technion.ac.il} \affiliation{Schulich Faculty of
Chemistry, Technion - Israel Institute of Technology, Haifa 32000,
Israel}
\begin{abstract}
We use shaped near-infrared (NIR) pulses to control the generation
of coherent broadband ultraviolet (UV) radiation in an atomic
resonance-mediated (2+1) three-photon excitation. Experimental and
theoretical results are presented for phase 
controlling the total emitted UV yield in atomic sodium (Na). Based
on our confirmed understanding, we present a new simple scheme for
producing shaped femtosecond pulses in the UV/VUV spectral range
using the control over atomic resonance-mediated generation of third
(or higher order) harmonic.
\end{abstract}

\maketitle


Controlling quantum matter dynamics with light is of great
scientific and technological importance. Optical coherent control is
aimed at directing the irradiated quantum system to desired final
state(s) by utilizing the coherent nature of light. The control over
the transition probabilities is obtained by manipulating the
interferences between various photo-induced state-to-state quantum
pathways. The broad coherent spectrum of femtosecond pulses provides
a wide coherent band of such pathways, thus presenting a unique
coherent control tool \cite{1,2,3,4,5}. The control knobs are the
amplitude, phase, and/or polarization of the various spectral
components of the pulse, manipulated by pulse shaping techniques \cite{6}.

Among the processes, over which such femtosecond control has been
very effective, are multiphoton processes
\cite{5,7,8,9,10,11,12,13,14,15,16,17,18,19,20,21,22}. Several of
the corresponding works involve controlling nonlinear optical
processes that lead to the emission of stimulated coherent radiation.
Examples include rational coherent over anti-stokes Raman scattering (CARS) \cite{8,18}
and "black-box" automatic adaptive control control over
the spectrum emitted in high-harmonic generation process \cite{23,24}.
Beyond its fundamental scientific interest, such control over the emission of
coherent stimulated radiation has applicative importance for nonlinear spectroscopy
and microscopy as well as for remote detection.
In addition, controlling the generation of short-wavelength coherent radiation can
provides means for producing shaped femtosecond pulses in the ultraviolet (UV) and
vacuum-ultraviolet (VUV) spectral regions. The availability of such shaped UV/VUV pulses
would greatly extend the variety of molecular systems that could be
coherently controlled, since most of the molecular electronic
transitions (either via a single- or multi-photon absorption) are at the UV/VUV frequencies.
The technical difficulty in UV/VUV pulse shaping is the low optical
transmission and low damage threshold at UV/VUV frequencies of the
materials used for the common near-infrared/visible (NIR/VIS) pulse
shaping devices. So far, there has only been a very small number of
works achieving UV pulse shaping \cite{25,26,27,28,29,30,31,32,33}
and all of them are based on converting NIR pulses to UV pulses
using nonlinear optical crystals, which limits the UV wavelength to
being longer than 250~nm.
One approach is direct shaping of the generated UV pulses using
devices that are tailored for specific spectral ranges
\cite{25,26,27,28,29,30,31}. Another approach is complicated
indirect shaping of the UV pulses by shaping the generating NIR
pulses \cite{32,33}.

Here, we demonstrate for the first time the use of shaped NIR pulses to control the generation
of coherent broadband UV radiation, i.e., a UV pulse, in an atomic resonance-mediated (2+1) three-photon excitation.
The resonance-mediated nature of the excitation provides high degree of control over the emitted UV radiation
as compared to non-resonant generation using nonlinear crystals.
The work is related to our previous study on controlling resonance-mediated (2+1) three-photon absorption
to a single final (real) state \cite{19}. The present excitation is of the same type, however it accesses
simultaneously and coherently a manifold of final ("virtual") states.
The model system is atomic sodium (Na), for which experimental and theoretical results are
presented for phase controlling the total emitted UV yield.
We also present a new simple scheme for producing shaped femtosecond pulses in the UV/VUV
spectral range based on controlling the atomic resonance-mediated
generation of third (or higher order) harmonic.
The emitted UV/VUV wavelength is determined by the wavelength of the driving shaped pulse and
by the state energies associate with the physical system under control.
Hence, in general, the proposed scheme can enable going down to short wavelengths that are
inaccessible with nonlinear optical crystals (see above).

\begin{figure} [htbp]
\includegraphics[scale=0.7]{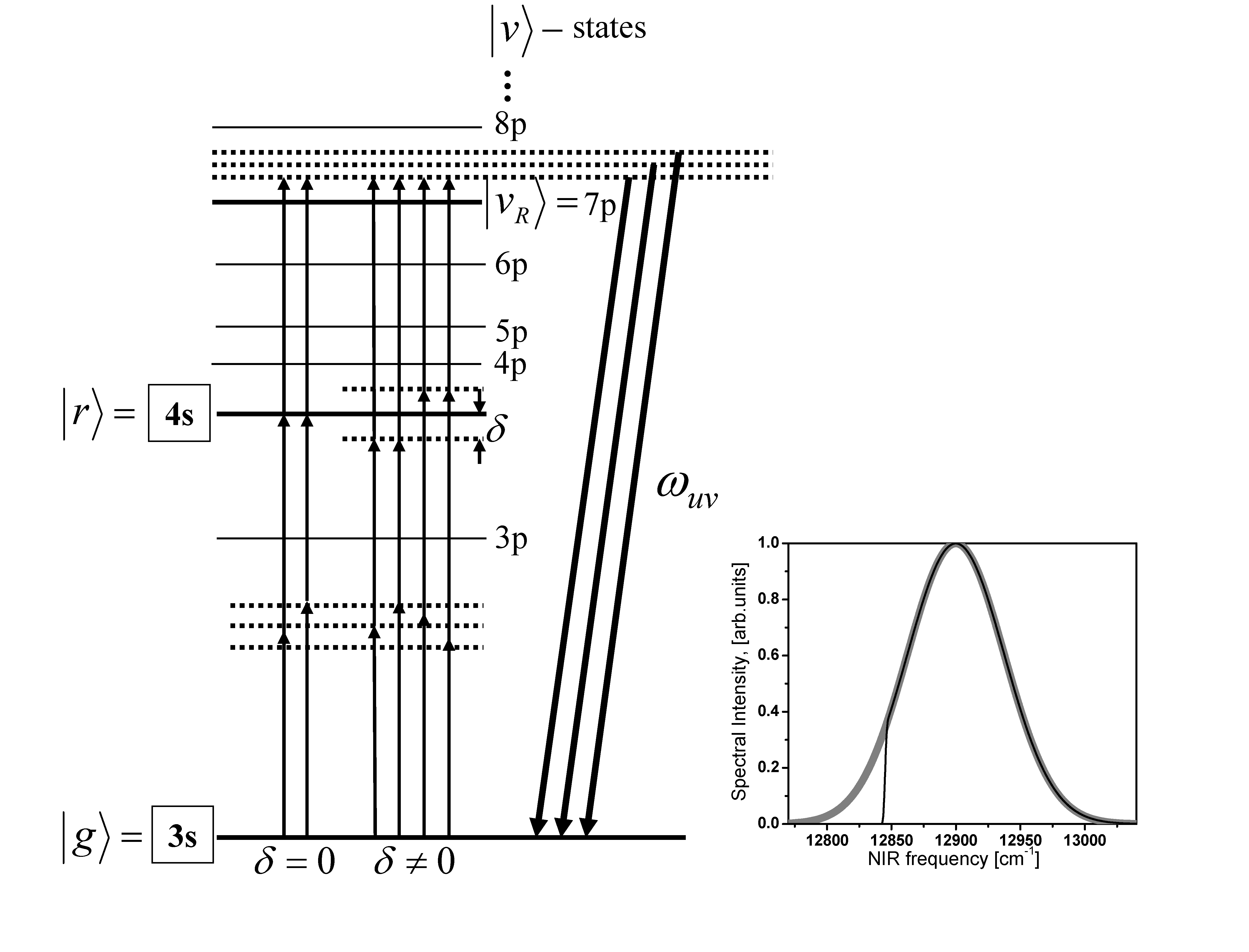}
\caption{Generation of coherent broadband UV emission via resonance-mediated (2+1) three-photon
excitation in Na. Several sets of three-photon
pathways that are on resonance ($\delta=0$) or near resonance
($\delta\neq0$) with $\left|r\right\rangle \equiv 4s$ are shown.
The inset shows the two excitation pulse spectra with (gray line) and without (black line)
a resonant access to $\left|v_R\right\rangle \equiv 7p$.}
\label{fig_1}
\end{figure}

We consider the atomic resonance-mediated (2+1) third-harmonic generation (THG) process
induced by a near-infrared (NIR) femtosecond pulse that is depicted in Fig.~\ref{fig_1}.
It involves an initial ground state $\left|g\right\rangle$ and an excited state $\left|r\right\rangle$
that are of one symmetry, and a manifold of excited states $\left|v_j\right\rangle$ that are of the other symmetry.
The NIR excitation pulse spectrum is such that all the $\left|g\right\rangle$-$\left|v_j\right\rangle$ and
$\left|r\right\rangle$-$\left|v_j\right\rangle$ couplings are non-resonant, except for the
$\left|r\right\rangle$-$\left|v_R\right\rangle$ coupling that, according to the considered case,
is either resonant or non-resonant.
Additionally, the excitation spectrum contains half the $\left|r\right\rangle$-$\left|g\right\rangle$
transition frequency ($\omega_{r,g}/2$). 
Hence, the irradiation with the NIR broadband pulse leads to a resonance-mediated three-photon excitation
from $\left|g\right\rangle$ to a broad range of final energies followed by a stimulated de-excitation back to
$\left|g\right\rangle$ emitting a coherent broadband UV radiation.
In the time-domain picture, the UV emission results from a time-dependent dipole moment induced by the NIR pulse.

Using 3$^{rd}$-order time-dependent perturbation theory, we have obtained the spectral 
UV field emitted at a UV frequency $\omega_{UV}$ to be given by 
\begin{eqnarray}
E_{TH}(\omega_{UV})&\propto&\mu^{2}_{r,g}[D^{(UV)}_{R}(\omega_{UV})
+ D^{(UV)}_{nonR}] A^{(2+1)}_{TH}(\omega_{UV}) \; ,
\label{eq1} \\
A^{(2+1)}_{TH}(\omega_{UV})&=&A^{(2+1)on-res}(\omega_{UV})+A^{(2+1)near-res}(\omega_{UV}) \; ,
\label{eq2} \\
A^{(2+1)on-res}(\omega_{UV})&=&i\pi E(\omega_{UV}-\omega_{r,g})A^{(2)}(\omega_{r,g}) \; ,
\label{eq3} \\
A^{(2+1)near-res}(\omega_{UV})&=&-\wp\int_{-\infty}^{\infty}\frac{1}{\delta}A^{(2)}(\omega_{r,g}-\delta)E(\omega_{UV}-\omega_{r,g}+\delta)d\delta \; ,
\label{eq4} \\
A^{(2)}(\Omega)&=&\int_{-\infty}^{\infty} E(\omega)E(\Omega-\omega)d\omega \; ,
\label{eq5}
\end{eqnarray}
where $E(\omega)\equiv\left|E(\omega)\right|\exp{[i\Phi(\omega)]}$ is the (NIR) spectral field of the excitation pulse,
with $\left|E(\omega)\right|$ and $\Phi(\omega)$ being, respectively, the spectral
amplitude and phase at frequency $\omega$.
For the (unshaped) transform-limited (TL) pulse, which is the shortest pulse for a given spectrum $\left|E(\omega)\right|$,
$\Phi(\omega)=0$ for any $\omega$.
The quantity $\mu^{2}_{r,g}$ is the
$\left|g\right\rangle$$\rightarrow$$\left|r\right\rangle$ effective
non-resonant two-photon excitation coupling, while
$D^{(UV)}_{R}(\omega_{UV})$ and $D^{(UV)}_{nonR}$ stand,
respectively, for the
$\left|r\right\rangle$$\rightarrow$$\left|v_{R}\right\rangle$$\rightarrow$$\left|g\right\rangle$
coupling via $\left|v_{R}\right\rangle$ and for the
$\left|r\right\rangle$$\rightarrow$$\left|v_{j}\right\rangle$$\rightarrow$$\left|g\right\rangle$
coupling via all the other $v_j$ states. They are given by
\begin{eqnarray}
D^{(UV)}_{R}(\omega_{UV})&=&\frac{\mu_{g,v_{R}}\mu_{v_{R},r}}{\omega_{v_{R},g}-\omega_{UV}+i\Gamma_{v_{R}}} \; ,
\label{eq6} \\
D^{(UV)}_{nonR}&=&\sum_{v_{j}\neq
v_{R}}^{\infty}\frac{\mu_{g,v_{j}}\mu_{v_{j},r}}{\omega_{v_{j},r}-\omega_0} \; ,
\label{eq7}
\end{eqnarray}
where $\mu_{m,n}$ and $\omega_{m,n}$ are the transition dipole moment and transition frequency between a pair of states,
and $\Gamma_{v_{r}}$ is the linewidth of $\left|v_{R}\right\rangle$.
The spectral intensity emitted at a UV frequency $\omega_{UV}$ is $I_{TH}(\omega_{UV})=\left|E_{TH}(\omega_{UV})\right|^{2}$.
The total UV yield is $Y_{TH}=\int_{-\infty}^{\infty}I_{TH}(\omega_{UV})d\omega_{UV}$.

As illustrated in Fig.~\ref{fig_1}, Eqs.~(\ref{eq1})-(\ref{eq7}) reflects the fact that 
the (complex) spectral field $E_{TH}(\omega_{UV})$ at each emitted UV frequency $\omega_{UV}$
results from the interferences between all the three-photon pathways
starting from $\left|g\right\rangle$ and reaching the final excitation energy that corresponds to $\omega_{UV}$.
Each such pathway is either on resonance or near resonance with the intermediate state $\left|r\right\rangle$,
having a corresponding detuning $\delta$. It involves a non-resonant absorption of
two photons with a two-photon transition frequency $\omega_{r,g}-\delta$
and the absorption of a third complementary photon of frequency $\omega_{UV}-(\omega_{r,g}-\delta)$.
The term $A^{(2+1)on-res}(\omega_{UV})$ [Eq.~(\ref{eq3})]
interferes all the on-resonant pathways ($\delta=0$),
while the term $A^{(2+1)near-res}(\omega_{UV})$ [Eq.~(\ref{eq4})] interferes all the
near-resonant pathways ($\delta\ne0$) with a 1/$\delta$ amplitude weighting.
The on-resonant pathways are excluded from $A^{(2+1)near-res}(\omega_{UV})$ by the
Caushy's principle value operator $\wp$.
Hence, in general, the emitted coherent broadband UV radiation is a shaped one, i.e.,
each $\omega_{UV}$ acquires its own amplitude and phase. One should note that, due to the
resonance-mediated nature of the excitation, this applies also for an excitation
with a transform-limited NIR pulse.


When the excitation pulse spectrum allows resonant access to
the $\left|v_{R}\right\rangle$ state by three-photon pathways,
the emitted UV spectrum contains the corresponding frequency $\omega_{UV}=\omega_{v_R,g}$.
Since the coupling component $D^{(UV)}_{R}(\omega_{UV})$ [Eq.~(\ref{eq6})] associated with $\left|v_{R}\right\rangle$
has a narrow response around $\omega_{UV}=\omega_{v_R,g}$,
with a magnitude that is much larger than the magnitude of
the other coupling component $D^{(UV)}_{nonR}$ [Eq.~(\ref{eq7})]
associated with all the other $\left|v\right\rangle$ states,
the emitted UV spectrum consists of a dominant narrowband part
centered around $\omega_{v_{R},g}$ that is superimposed on
a broadband part of much smaller magnitude.
The corresponding total UV yield is then proportional to the spectral intensity
at $\omega_{v_R,g}$, i.e.,
$Y_{TH} \propto I_{TH}(\omega_{v_R,g}) \propto |A^{(2+1)}_{TH}(\omega_{v_{R},g})|^2$.
This is actually proportional to the population excited to $\left|v_{R}\right\rangle$ by the
femtosecond pulse \cite{19}.
When there is no resonant access to $\left|v_{R}\right\rangle$,
the $D^{(UV)}_{R}(\omega_{UV})$ coupling is effectively non-resonant
and independent of the emitted $\omega_{UV}$, similar to $D^{(UV)}_{nonR}$.
The UV emission spectrum is then broadband with no narrowband component.

The above excitation scheme is physically realized here with atomic sodium (Na) (see Fig.~\ref{fig_1}),
having the $3s$ ground state as $\left|g\right\rangle$, the $4s$ state as $\left|r\right\rangle$,
and the manifold of $p$-states as $\left|v_j\right\rangle$ with $7p$ as $\left|v_R\right\rangle$.
The transition frequency $\omega_{r,g} \equiv \omega_{4s,3s} = 25740$~cm$^{-1}$ corresponds to two 777-nm photons and the
frequency $\omega_{v_R,r} \equiv \omega_{7p,4s} = 12801$~cm$^{-1}$ corresponds to a 781.2-nm photon.
The sodium is irradiated with phase-shaped linearly-polarized femtosecond pulses having a Gaussian
intensity NIR spectrum centered around 775.2~nm with 5.3-nm (FWHM) bandwidth ($\sim$165-fs TL duration).
Experimentally, a sodium vapor in a heated cell is irradiated with such laser pulses, after
they undergo shaping in an optical setup incorporating a pixelated
liquid-crystal spatial light phase modulator \cite{6}.
The effective spectral shaping resolution is $\delta\omega_{shaping}$=2.05~cm$^{-1}$ 
per pixel.
The peak intensity of the TL pulse is about 10$^{9}$~W/cm$^{2}$. 
Following the interaction with the NIR pulse, the coherent UV radiation emitted in the forward direction
is separated from the NIR excitation pulse using a proper optical filter and is 
measured by a spectrometer coupled to a camera system.


As a first control study on resonance-mediated UV generation, this
work focuses on controlling the total UV yield $Y_{TH}$ with shaped
pulses having spectral phase patterns of a $\pi$-step at variable
position $\omega_{step}$. As previously shown, this family of pulse
shapes is highly efficient in controlling two-photon absorption
\cite{10} and resonance-mediated (2+1) three-photon absorption
\cite{19}. The phase control over the UV yield is studied in two
cases: when the excitation spectrum allows access to the $7p$ state
(via various three-photon pathways) and when it does not (see
above). It is implemented by unblocking or blocking the
low-frequency end of the excitation pulse spectrum (see inset of
Fig.~\ref{fig_1}).

\begin{figure} [htbp]
\includegraphics[scale=0.7]{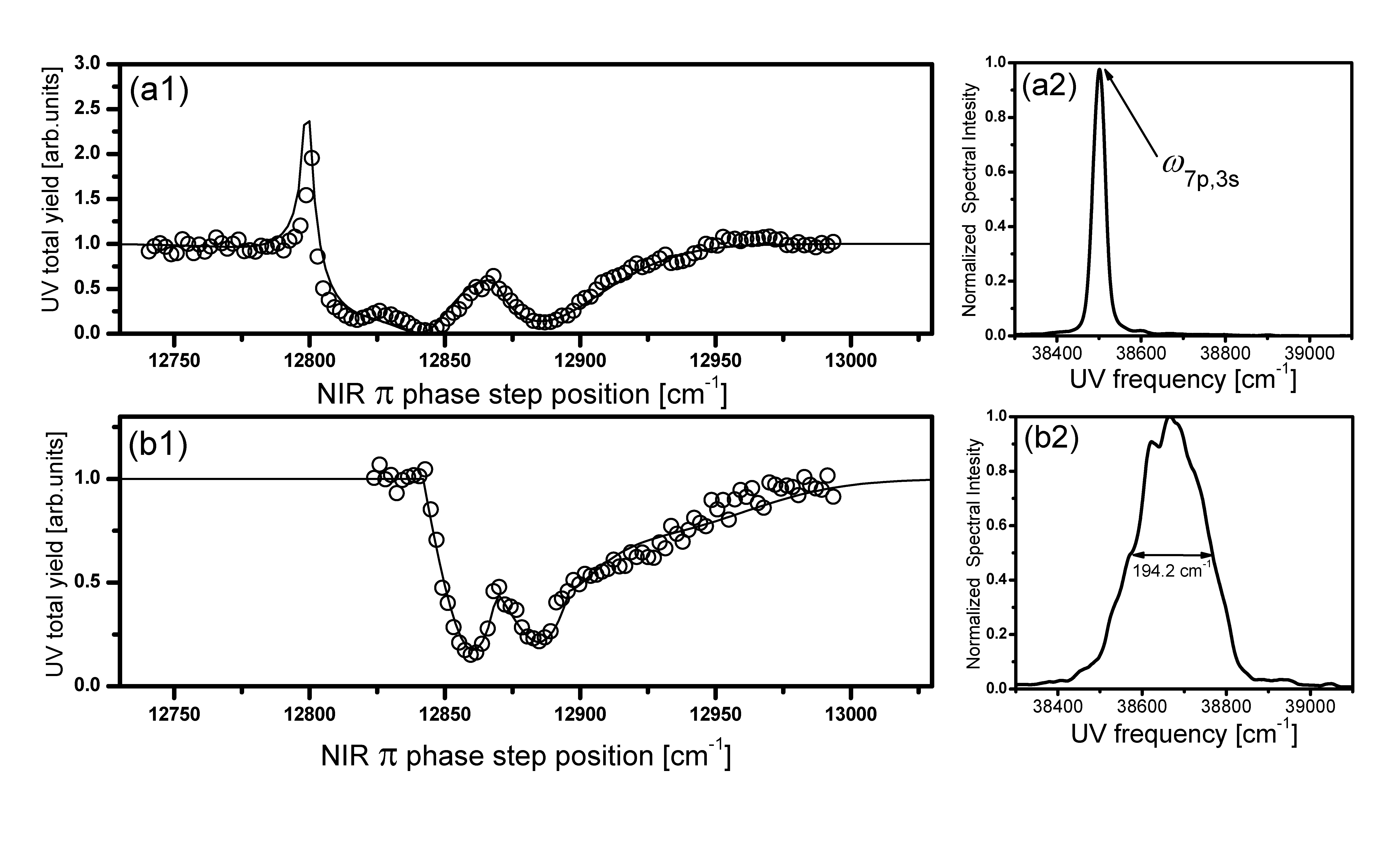}
\caption{Experimental (circles) and theoretical (solid lines)
results for the total UV total yield generated by shaped pulses with
spectral $\pi$ phase step, when the excitation pulse spectrum (a1) allows
and (b1) blocks the resonant access to $7p$ (shown in Fig.~\ref{fig_1}).
The UV yield is shown as a function of the position of the $\pi$ phase step.
The traces are normalized by the yield generated by the corresponding transform-limited (TL) pulse.
The UV spectra generated by the TL pulses of the two cases are shown in panels (a2) and (b2).}
\label{fig_2}
\end{figure}

Figure~\ref{fig_2} presents experimental (circles) and theoretical (lines) results
for the total UV yield $Y_{TH}$ as a function of the $\pi$-step position $\omega_{step}$.
The results for the cases of the $7p$ state being accessible or inaccessible
are shown, respectively, in Figs.~\ref{fig_2}(a1) and \ref{fig_2}(b1).
Each trace is normalized by $Y_{TH}$ induced by the corresponding TL pulse.
The theoretical results are calculated numerically using Eqs.(\ref{eq1})-(\ref{eq7}), using a grid
with a bin size equal to the experimental shaping resolution. 
As can be seen, there is an excellent agreement between the experimental and theoretical results,
confirming our theoretical description and understanding.
The UV spectra measured in the two cases with a TL excitation are also shown in Fig.~\ref{fig_2}.
As seen and explained above, an access to the $7p$ state leads to a
UV spectrum that is dominated by a strong narrowband part around
$\omega_{UV} = \omega_{7p,3s}$, and when the access to the $7p$
state is blocked the UV spectrum is purely broadband
[Fig.~\ref{fig_2}(a1)]. Here, it is of 194.2-cm$^{-1}$ (FWHM)
bandwidth.

\begin{figure} [htbp]
\includegraphics[scale=0.7]{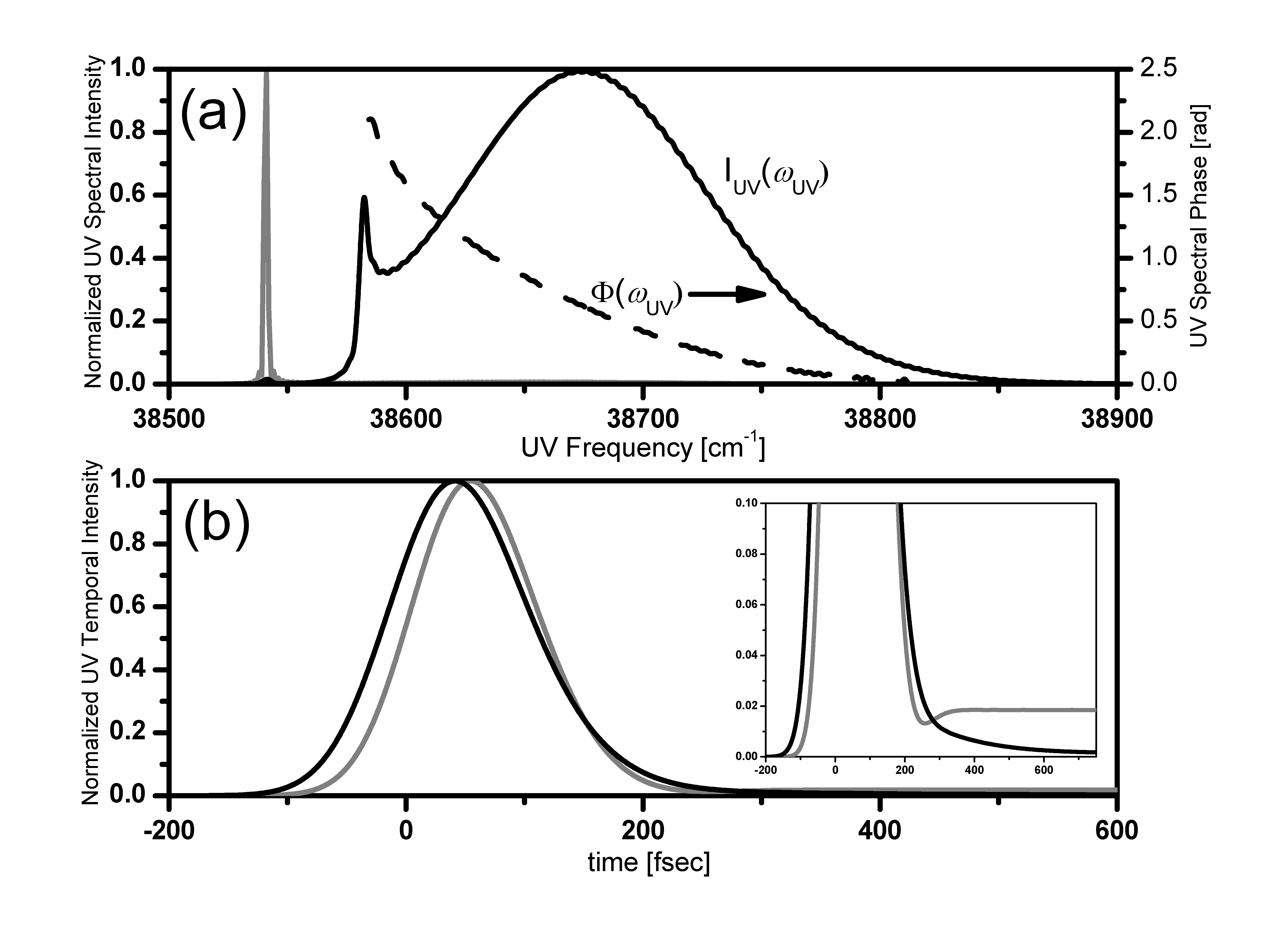}
\caption{Theoretical results of the UV pulses obtained by the TL excitation pulses.
(a) Calculated UV spectrum in the narrowband case
when the $7p$ state is accessible (gray) and in the broadband case
when the $7p$ state is inaccessible (black). The former includes
both intensity and phase information.
(b) Calculated temporal UV pulses in the narrowband case (gray)
and in the broadband case (black).
The calculations are direct time-integration of the Schr\"{o}dinger equation.}
\label{fig_3}
\end{figure}

Considering first the case when $7p$ is resonantly accessed. As explained above, the total yield $Y_{TH}$
in this case is proportional to the population excited to the $7p$ state $P_{7p}$. Indeed, the TL-normalized
$\pi$-trace shown in Fig.~\ref{fig_2}(a1) for $Y_{TH}$ reproduces the one measured for $P_{7p}$
with similar NIR excitation \cite{19}.
The total UV yield is experimentally controlled from 3\% to about
200\% of the yield induced by the TL pulse The strong enhancement
occurs when $\omega_{step}=\omega_{7p,4s}$=12801~cm$^{-1}$. As
previously identified for the resonance-mediated three-photon
absorption \cite{19}, it originates from a change in the nature of
the interferences between the positively-detuned ($\delta$$>$0) and
negatively-detuned ($\delta$$<$0) near-resonant $3s$-$7p$
three-photon pathways. With the TL pulse they are destructive, while
with a $\pi$-step at $\omega_{7p,4s}$ they are constructive.
The physical reason for this proportionality between $Y_{TH}$ and $P_{7p}$ is the coherent superposition
of the $3s$ and $7p$ states that is created by the excitation and survives also after the pulse is over,
leading to a long-lived time-dependent dipole moment.
This dipole moment induces the UV emission at frequency $\omega_{7p,3s}$.
The lifetime of the $3s$-$7p$ superposition is determined by the
experimental decoherence time, estimated in our case to be a few nanoseconds.
Hence, in this case, the overall result is a ultrashort UV pulse of small integrated energy,
followed by a very long quasi-monochromatic radiation at $\omega_{7p,3s}$ with large integrated energy.
This is verified by the theoretical results presented in Fig.~\ref{fig_3} that we have calculated
numerically by the direct time-integration of the Schr\"{o}dinger Equation.
Fig.~\ref{fig_3}(a) shows (in gray line) the calculated narrowband UV spectrum for a TL
excitation pulse with a spectral access to $7p$, and Fig.~\ref{fig_3}(b) shows
the corresponding calculated temporal UV emission (gray line) composed of a 122.5-fsec pulse
followed by a continuous radiation.
The temporal UV intensity reflects the fact that the narrowband UV spectrum is
actually superimposed on a (much weaker) broadband component.

In the other case when the $7p$ is inaccessible [Fig.~\ref{fig_2}(b)] the total UV yield $Y_{TH}$
is not dominated by any single frequency.
Similar to the above enhancement with $\omega_{step}=\omega_{7p,4s}$ when the $7p$ is accessed,
a $\pi$-step at a position $\omega_{step}$ enhances here the
UV amplitude at frequencies around $\omega_{UV}=\omega_{step}+\omega_{4s,3s}$, while
it affects the amplitudes of other frequencies in a complicated way (leading to attenuation or enhancement)
according to the above theoretical description.
Thus, no enhancement of the total UV yield $Y_{TH}$ beyond the TL level
is observed in Fig.~\ref{fig_2}(b1) for any $\omega_{step}$.
The corresponding control is from about 10\% to 100\% of the UV yield induced by the TL pulse.
Under our experimental conditions, the integrated energy of the
broadband UV emission is about 3\% of the NIR pulse energy (here, 5~$\mu$J) for a TL excitation.
The results from the numerical integration of the Schr\"{o}dinger Equation for the TL excitation
are shown in Fig.~\ref{fig_3}.
Fig.~\ref{fig_3}(a) shows the intensity and phase of the broadband UV emission.
The former is in agreement with the measurement shown in Fig.~\ref{fig_2}(b2).
Fig.~\ref{fig_3}(a) shows the calculated temporal UV pulse having a 134.7-fsec duration,
and, as expected, is not followed by continuous UV emission.
As can be seen the varying UV spectral phase, this UV pulse is not a transform-limited one;
the TL duration corresponding to the calculated Uv spectrum is 88~fsec.
It also worth noting that the emitted UV pulse is time delayed with respect to the driving NIR pulse
(its peak is not at $t$=0) due to the resonance-mediated nature of the excitation.

\begin{figure} [!h] 
\includegraphics[scale=0.5]{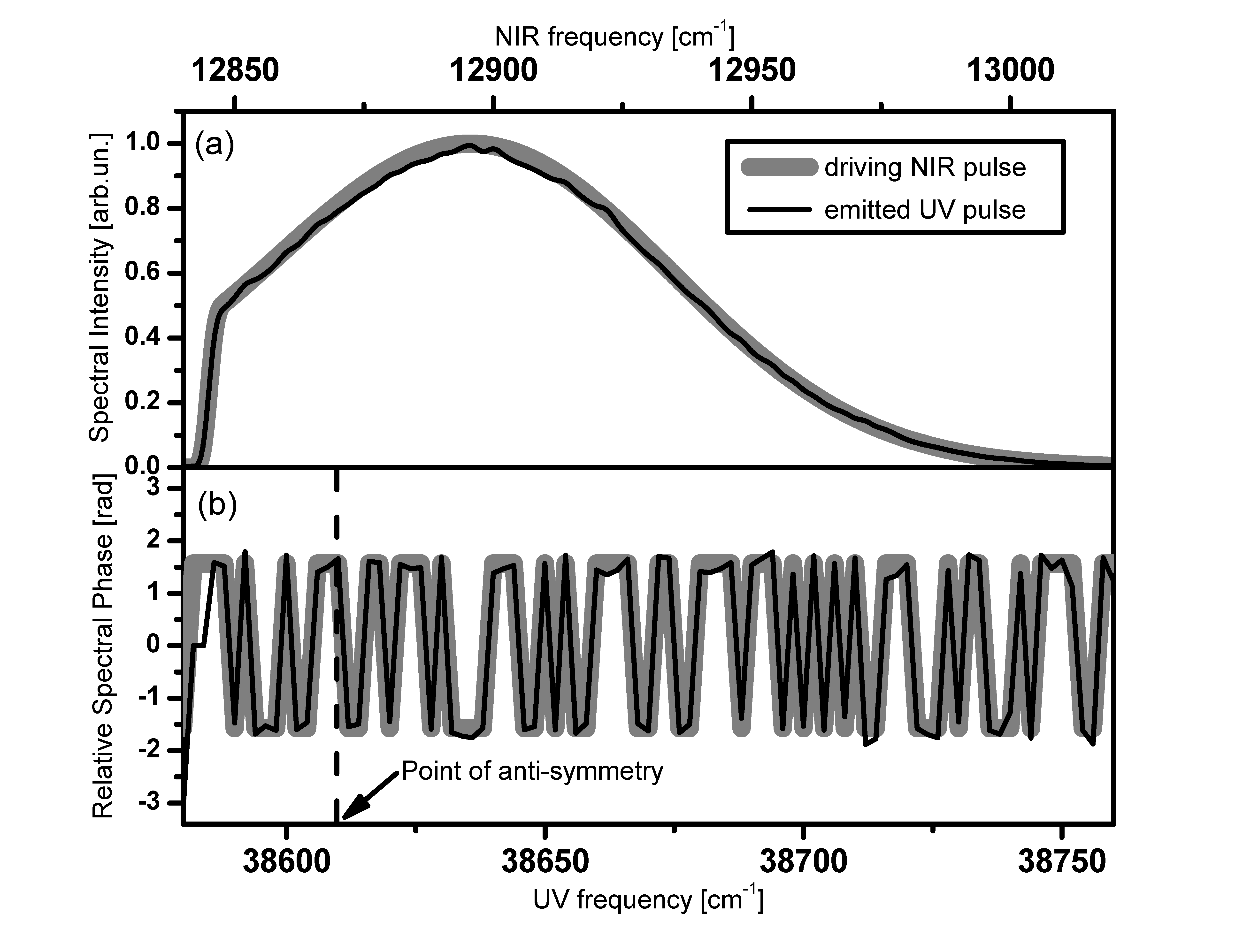}
\caption{Numerical example of the direct spectral amplitude and phase transfer from
the NIR excitation pulse to the UV pulse, when the NIR spectral phase pattern is anti-symmetric
around $\omega_{4s,3s}/2$.
(a) The NIR excitation spectrum (gray) and the generated UV spectrum (black).(b) The spectral
phase of the NIR excitation pulse (gray) and of the emitted UV pulse (black).} \label{fig_4}
\end{figure}

Continue considering the case without a resonant access to $\left|v_R\right\rangle$ ($7p$),
as can be seen from Eqs.~(\ref{eq1})-(\ref{eq7}), in general,
the shape of the emitted UV spectral field $E_{TH}(\omega_{UV}$ has a rather
complicated dependence on the shape of the exciting NIR field $E_{NIR}(\omega_{NIR}) \equiv E(\omega)$,
with both on-resonant and near-resonant amplitude contributions.
However, this dependence may be significantly simplified if one applies a NIR spectral
phase patterns that attenuate the near-resonance component $A^{(2+1)near-res}(\omega_{UV})$ [Eq.~(\ref{eq4}]
relative to the on-resonant component $A^{(2+1)on-res}(\omega_{UV})$ (Eq.~(\ref{eq3})].
The simplicity originates from the fact that the on-resonant term
simply transfers the NIR spectral amplitudes (up to a
proportionality constant) and phases (up to a global phase) to the
UV spectral field with a one-to-one frequency mapping of
$E_{UV}(\omega_{UV})\propto
E_{NIR}(\omega_{NIR}=\omega_{UV}-\omega_{r,g})$,
Such attenuation of the near-resonant component relative to the on-resonant one is actually
obtained by most of the spectral phase patterns that are anti-symmetric around
$\omega_{r,g}/2$ ($\omega_{4s,3s}/2$), i.e., $\Phi(\omega)=-\Phi(\omega_{r,g}-\omega)$.
A shaped NIR pulse with such an anti-symmetric spectral phase induces fully constructive
interferences only among all the three-photon pathways that are on-resonant
with $\left|r\right\rangle$ ($4s$),
thus keeping $\left|A^{(2+1)on-res}(\omega_{UV})\right|$ on its maximal value for $\omega_{UV}$.
while generally attenuating $\left|A^{(2+1)on-res}(\omega_{UV})\right|$.
The case of each individual anti-symmetric phase pattern
can easily be predicated numerically based on Eqs.~(\ref{eq1})-(\ref{eq7}).
To illustrate this basic idea, which can be a basis for a simple scheme for producing shaped
UV femtosecond pulses, a numerical example is given in Fig.~\ref{fig_4}.
We have calculated numerically the UV spectral field generated by irradiating Na with
a NIR pulse having the randomly selected anti-symmetric spectral phase pattern shown
in Fig.~\ref{fig_4}(b) (gray line; top x-axis scale).
The NIR spectrum is shown in Fig.~\ref{fig_4}(a) (gray line; top x-axis scale).
The resulting UV spectral intensity and phase are shown, respectively,
in Fig.~\ref{fig_4}(a) and (b) (black lines, bottom x-axis scale).
As can be seen, 
there is almost-perfect direct one-to-one amplitude and phase
transfer from the NIR to the UV with a frequency shift of $\omega_{UV} - \omega_{NIR} = \omega_{4s,3s}$.
In other words, the spectral shape of the excitation NIR pulse is directly imprinted on the emitted UV pulse.
The above basic idea can be further extended.
For example, when an anti-symmetric UV spectral phase is undesired, one can filter out parts of the UV spectrum
leaving only the parts of interest.
Also, one can, for example, use the part of the NIR spectrum around $\omega_{r,g}/2$
to attenuate the near-resonant component with a proper anti-symmetric phase, while encoding in
the desired phases in the other NIR spectral parts. Then, if needed, filter out the UV part resulting from
the "control" part.
Last, it worth noting that, with NIR excitation of the Na atom, such a scheme can allow producing shaped UV pulses
of up to about 500-600~cm$^{-1}$ bandwidth ($\sim$20~fsec TL duration).

In summary, first experimental and theoretical studies of phase controlling resonance-mediated
(2+1) generation of coherent broadband UV emission by shaped femtosecond pulses have been presented here.
With proper NIR excitation spectrum, the UV emission is of broad spectrum corresponding to a femtosecond UV pulse.
Based on our confirmed understanding,
we have also presented a new simple scheme for producing shaped UV/VUV pulses
using the control over atomic resonance-mediated generation of third (or higher order) harmonic.
The emitted UV/VUV wavelength is determined by the wavelength of the driving shaped pulse and
by the state energies associate with the physical system under control.
Hence, in general, the proposed scheme can enable going down to short wavelengths that are
inaccessible with other UV shaping techniques.
The availability of such a scheme would greatly extend the variety of molecules to be coherently controlled.

This research was supported by The Israel Science Foundation (grant No.~127/02),
by The James Franck Program in Laser Matter Interaction,
and by The Technion's Fund for The Promotion of Research.



\newpage

\begin{thebibliography}{9999}

\bibitem{1} D. J. Tannor, R. Kosloff,
and S. A. Rice, J. Chem. Phys. $\boldsymbol{85}$, 5805 (1986).

\bibitem{2} M. Shapiro and P. Brumer, {\it Principles of the quantum control of molecular
processes}(Wiley, New Jersey, 2003).

\bibitem{3} W. S. Warren, H. Rabitz, and D. Mahleh, Science $\boldsymbol{259}$, 1581 (1993).

\bibitem{4} H. Rabitz, R. de Vivie-Riedle, M. Motzkus, and K. Kompa, Science
$\boldsymbol{288}$, 824 (2000).

\bibitem{5} M. Dantus and V. V. Lozovoy,
Chem. Rev. $\boldsymbol{104}$, 1813 (2004);

\bibitem{6} A. M. Weiner, Rev. Sci. Inst. $\boldsymbol{71}$, 1929 (2000);
T. Brixner and G. Gerber, Opt. Lett. $\boldsymbol{26}$, 557 (2001);
T. Brixner {\it et al.}, Appl. Phys. B $\boldsymbol{74}$, S133
(2002).

\bibitem{7} N. Dudovich, D. Oron, and Y. Silberberg,
Phy. Rev. Lett. $\boldsymbol{92}$, 103003 (2004).

\bibitem{8} D. Oron {\it et al.}, Phys. Rev. A $\boldsymbol{65}$, 043408 (2002);
N. Dudovich, D. Oron, and Y. Silberberg, Nature (London)
$\boldsymbol{418}$, 512 (2002); J. Chem. Phys. $\boldsymbol{118}$,
9208 (2003).

\bibitem{9} M. Wollenhaupt {\it et al.}, Phys. Rev. A $\boldsymbol{68}$, 015401 (2003);
Chem. Phys. Lett. $\boldsymbol{419}$, 184 (2006).


\bibitem{10} D. Meshulach and Y. Silberberg,
Nature (London) $\boldsymbol{396}$, 239 (1998); Phys. Rev. A
$\boldsymbol{60}$, 1287 (1999).

\bibitem{11} K. A. Walowicz {\it et al.},
J. Phys. Chem. A $\boldsymbol{106}$, 9369 (2002); V. V. Lozovoy {\it
et al.}, J. Chem. Phys. $\boldsymbol{118}$, 3187 (2003).

\bibitem{12} A. Pr$\ddot{a}$kelt {\it et al.},
Phys. Rev. A $\boldsymbol{70}$, 063407 (2004).   

\bibitem{13} N. Dudovich {\it et al.},
Phys. Rev. Lett. $\boldsymbol{86}$, 47 (2001).

\bibitem{14} B. Chatel, J. Degert, and B. Girard,
Phys. Rev. A $\boldsymbol{70}$, 053414 (2004).

\bibitem{15}
P. Panek and A. Becker, Phys. Rev. A $\boldsymbol{74}$, 023408
(2006).

\bibitem{16}
E. Gershgoren {\it et al.}, Opt. Lett. $\boldsymbol{28}$, 361
(2003).

\bibitem{17}
H. U. Stauffer {\it et al.}, J. Chem. Phys. $\boldsymbol{116}$, 946
(2002); X. Dai, E. W. Lerch, and S. R. Leone, Phys. Rev. A
$\boldsymbol{73}$, 023404 (2006).

\bibitem{18}
S. Lim, A. G. Caster, and S. R. Leone, Phys. Rev. A
$\boldsymbol{72}$, 041803 (2005);

\bibitem{19} A. Gandman, L. Chuntonov, L. Rybak, and Z. Amitay,
Phys. Rev. A $\boldsymbol{75}$, 031401 (R) (2007); Phys. Rev. A, to
be published (http://arxiv.org/abs/0709.0601).

\bibitem{20}
L. Chuntonov, L. Rybak, A. Gandman, and Z. Amitay,
http://arxiv.org/abs/arXiv:0709.0486;
http://arxiv.org/abs/arXiv:0709.0615.

\bibitem{21} N. Dudovich {\it et al.},
Phys. Rev. Lett. $\boldsymbol{94}$, 083002 (2005).

\bibitem{22} C. Trallero-Herrero {\it et al.},
Phys. Rev. Lett. $\boldsymbol{96}$, 063603 (2006).

\bibitem{23} T. Pfeifer, D. Walter, C. Winterfeldt, C. Spielmann G. Gerber, Appl.
Phys. B $\boldsymbol{80}$, 277 (2005)

\bibitem{24} R. Bartels, S. Backus, E. Zeek, L. Misoguti, G. Vdovin, I.P. Christov, M.M. Murnane, H.C. Kapteyn, Nature
$\boldsymbol{406}$, 164 (2000)

\bibitem{25} B.J. Pearson, T.C. Weinacht Opt. Exp. $\boldsymbol{15}$, 4385 (2007)

\bibitem{26} M. Hacker, G. Stobrawa, R. Sauerbrey, T. Backup, M. Motzkus,
M. Wildenhaim, A. Gehner Appl. Phys. B. $\boldsymbol{76}$, 711
(2003)

\bibitem{27} G. Strorawa, M. Haker, T. Feurer, C. Zeidler, M. Motzkus, F.
Reichel, Appl. Phys. B $\boldsymbol{72}$, 627 (2001).

\bibitem{28} G. Schriever, S. Lochbrunner, M. Optiz, E. Riedle Opt.
Lett. $\boldsymbol{31}$, 543 (2006)

\bibitem{29} M. Hacker, R. Netz, M. Roth, G. Stobrawa, T. Feurer, and R.
Sauerbrey, Appl. Phys. B $boldsymbol{73}$, 273 (2001)

\bibitem{30} H. Wang and A. M. Weiner, IEEE J.
Quantum Electron. $boldsymbol{40}$, 937 (2004)

\bibitem{31} Y. Nabekawa and K. Midorikawa, Appl. Phys. B $boldsymbol{78}$,
569 (2004).


\bibitem{32} S. Shimizu, Y. Nabekawa, M. Obara, and K. Midorikawa, Opt.
Express $\boldsymbol{13}$, 6345 (2005).

\bibitem{33} M. Roth, M. Mehendale, A. Bartelt, H. Rabitz, Appl. Phys. B.
$\boldsymbol{80}$ 441 (2005)


\end{thebibliography}
\end{document}